\documentclass[10pt,twocolumn,letterpaper]{article}
\usepackage{graphicx}
\usepackage{subfigure}
\usepackage{pstool}
\usepackage{textcomp}

\newcommand{\myfigurewide}[3]
{\begin{figure*}[!tb]\begin{center}#2\caption{#3}\label{#1}\end{center}\end{figure*}}

\newcommand{\mygraphfigurewide}[4]{\myfigurewide{#1}{\psfragfig[#2]{#3}}{#4}}

\newcommand{\myfigure}[3]
{\begin{figure}[!tb]\begin{center}#2\caption{#3}\label{#1}\end{center}\end{figure}}

\newcommand{\mygraphfigure}[4]{\myfigure{#1}{\psfragfig[#2]{#3}}{#4}}

\newcommand{\mygraphsubfigure}[4]
{\subfigure[#4]{\hspace{0.5cm}\psfragfig[#2]{#3}\label{#1}\hspace{0.5cm}}}

\newcommand{\tensv}{\vec{T}}
\newcommand{\tens}{T}
\newcommand{\pos}{\vec{r}}
\newcommand{\s}{s}
\newcommand{\area}{A}
\newcommand{\stress}{\sigma_0}
\newcommand{\nablav}{\vec{\nabla}}
\newcommand{\density}{\rho}

\newcommand{\pot}{V}
\newcommand{\deltaV}{\Delta \pot}
\newcommand{\diff}[1]{\mathrm{d} #1}
\newcommand{\der}[2]{\frac{\diff{#1}}{\diff{#2}}}
\newcommand{\dder}[2]{\frac{\mathrm{d}^2 #1}{\diff{#2}^2}}

\begin{document}  
\psfragscanon


\title{Non-Equatorial Uniform-Stress Space Elevators\footnote{\copyright
2004 Institute for Scientific Research, Inc.}
\footnote{This paper was first published in \textit{Proc. of the 3rd
International Space Elevator Conference, June 2004}, reprinted on arXiv by
permission of Bradley Edwards former director at ISR.}}

\author{Blaise Gassend\footnote{The author may be contacted at
gassend@alum.mit.edu.}}
\date{}
\maketitle
\thispagestyle{empty}

\begin{abstract}

Non-equatorial space elevators are interesting as they give more freedom
for anchor location, avoid the highly occupied geosynchronous orbit and the
areas of highest radiation in the Van Allen belts. 
We review the static equation for a uniform-stress tether, and use
it to study the tapering of a uniform-stress tether 
in a general potential field. We then focus on a rotating coulomb potential
and study the range of latitudes that are allowed for a uniform-stress
space elevator. Finally, we look at a few practical issues that are raised
by non-equatorial space elevators.

\end{abstract}

\section*{Introduction}

\begin{table}[!t]
\begin{center}
\begin{tabular}{|c|p{6cm}|}
\hline
Symbol & Description \\
\hline
\hline
$s$ & Curvilinear coordinate along the tether. \\
\hline
$\vec{r}$ & Position vector of a point on the tether. \\
\hline
$r$ & Distance from the center of the planet. \\
\hline
$r_\perp$ & Distance from the rotation axis of the planet. \\
\hline
$r_s$ & Distance from the center of the planet to the synchronous
                       altitude. \\
\hline

$\rho$ & Density of tether material under stress. \\
\hline
$\sigma_0$ & Target stress in tether. \\
\hline
$A$ & Cross-sectional area of tether. \\
\hline
$\vec{T}$ & Tension applied by the top part of the tether to the bottom part. \\
\hline
$m$ & Mass of the counterweight. \\
\hline

$V$ & Potential field the tether is placed in. \\
\hline

$\theta$ & Angle between the equatorial plane and the position vector
             $\vec{r}$. \\
\hline
$\phi$ & Angle between the equatorial plane and the tangent to the
           tether. \\
\hline
$\psi$ & Angle between the tangent to the tether and the position vector
           $\vec{r}$ \\
\hline
$\hat{e}_\phi$ & Unit vector in the direction of increasing $\phi$. \\
\hline

$G$ & Gravitational constant. \\
\hline
$M_p$ & Mass of the planet. \\
\hline
$\Omega$ & Angular velocity of planet
		rotation. \\
\hline
$V_0$ & Characteristic specific energy of the rotating
         Coulomb potential field. \\
\hline
$\Delta V$ & Difference in potential between the base of the tether and 
		the point where the potential is greatest. \\
\hline
$\vec{\tilde{g}}$ & Combined gravitational and centrifugal field, in normalized
	form.\\ 
\hline
$\alpha$ & Tether shape factor. \\
\hline
$P/M$ & Payload to tether mass ratio. \\
\hline
$\left(\vec{v}\right)_\perp$ & Part of some vector $\vec{v}$ that is normal to the 
 	tether. \\
\hline
$\tilde{d}$ & Distance $d$ in units of $r_s$. \\
\hline
$\breve{d}$ & Distance $d$ in units of $\alpha r_s$. \\
\hline
$x_0$ & The value of variable $x$ at the anchor point (except $V_0$ and
          $\sigma_0$). \\
\hline
\end{tabular}
\end{center}
\caption{Notation that is used in this paper.}
\label{tab:notation}
\end{table}

\begin{table}[!t]
\begin{center}
\begin{tabular}{|c|c|p{12cm}|}
\hline
Symbol & Typical Value \\
\hline
\hline
$G$ & $ 6.67 \cdot 10^{-11}$ SI \\
\hline
$M_p$ & $5.98 \cdot 10^{24}$ kg \\
\hline
$\Omega$ & $7.29 \cdot 10^{-5}$ rad/s \\
\hline
$V_0$ & $9.46 \cdot 10^6$ J/kg \\
\hline
$r_s$ & $42.2 \cdot 10^6$ m \\
\hline
$r_0$ & $6.38 \cdot 10^6$ m \\
\hline
$\tilde{r}_0$ & 0.151 \\
\hline
$\rho$ & 1300 kg/m$^3$ \\
\hline
$\sigma_0$ & $65\cdot10^9$ N/m$^2$ \\
\hline
$\alpha$ & 0.189 \\
\hline
\end{tabular}
\end{center}
\caption{Typical values for the Earth and Edwards' tether 
parameters~\cite{edwards2002}. When
nothing is specified, these values are used for examples.}
\label{tab:typicalvalues} 
\end{table}

A space elevator is a very long tether. One end is attached to an anchor
station on the surface of a planet. The other end is attached to a
counterweight located beyond the planet's synchronous altitude. Past that
altitude centrifugal force due to the planet's rotation exceeds the
gravitational force, so the counterweight is pulled away from the planet.
Thus, the tether is kept in tension between anchor and counterweight.
Payloads can be sent into space by climbing the tether, much more
efficiently than with rockets.

The space elevator concept was first proposed in Russian
\cite{artsutanov60,lvov67} by Artsutanov, and later introduced in English by 
Pearson \cite{pearson75}.  Science fiction writers \cite{clarke79} made the
idea popular. But materials with the necessary strength to weight ratio
seemed unlikely. And the proposed elevators were much too heavy for
foreseeable launch technologies. This changed with the arrival of carbon 
nanotubes, and the
proposal by Edwards~\cite{edwards2002} of a space elevator concept that
would only require a few tons of material to be lifted by conventional
launch means.

To date, studies have considered that the space elevator would be anchored on
the equator, as that is where the equilibrium of the tether is easiest to
understand. Clarke~\cite{clarke79} even considered that the elevator would
have to be placed on the equator at a local minimum of the Earth's
geopotential. In fact, there is no reason for such limitations,
and non-equatorial space elevators can present a number of advantages:

\begin{itemize}

\item There is increased flexibility in the selection of the anchor
location.

\item The tether does not have to go through the heavily occupied (geo)synchronous 
orbit.

\item The tether avoids the areas of most intense radiation of the
Van Allen belts \cite{jorgensen2004}. This is particularly important when considering the
transportation of humans.

\item The tether can avoid the Martian moons (in the case of a Martian
elevator).

\end{itemize}

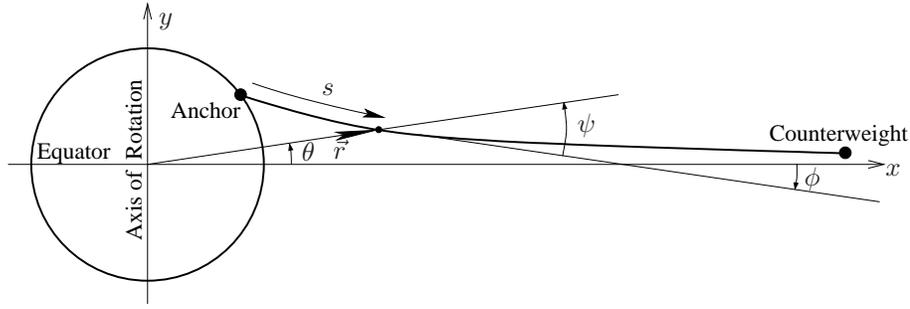
\mygraphfigurewide{fig:nomenclature}{width=5.25in}{notation}{A non-equatorial
space elevator, and some of the angles that are used to characterize it.}

Figure~\ref{fig:nomenclature} shows a diagram of a non-equatorial space
elevator. Three forces hold it in equilibrium: gravity, centrifugal force, and 
the tension at the anchor. The major difference with equatorial elevators
is that the elevator is located entirely on one side of the equatorial
plane. Therefore, gravity tends to pull the elevator towards the equatorial plane.
This tendency is countered by the force applied by the anchor, 
allowing the elevator to be in equilibrium.

In this paper we will be considering uniform-stress space-elevators. A
uniform-stress tether is a tether in which the cross-section is varied
(tapered) in order to keep the stress in the tether uniform.  Space
elevators are generally designed to have uniform stress as this maximizes
the ratio of payload mass to elevator mass.

To understand off-equator space elevators, we will first review the static
equations that any uniform-stress tether must satisfy, in Section~\ref{sec:equations}.
Then we will apply these equations to the relevant case of a rotating
Coulomb potential in Section~\ref{sec:coulomb}. In
Section~\ref{sec:maxlatt} we will study the problem of determining the maximum
latitude a space elevator can start from. Finally,
Section~\ref{sec:practical} covers a few practical concerns that are
specific to non-equatorial space elevators.

All the notation that is used in this paper can be found in
Table~\ref{tab:notation}. Examples will often use values that are typical
for Earth and the tethers considered by Edwards~\cite{edwards2002}.
Table~\ref{tab:typicalvalues} summarizes these values.

\section{Equations for a Static Uniform Stress Tether}
\label{sec:equations}

First we introduce the equations for a static uniform-stress tether in a
potential $\pot$.

\begin{eqnarray}
\label{eq:noshear}
  \der{\pos}{\s} & = & \frac{\tensv}{\tens} \\
\label{eq:tens}
  \der{\tensv}{\s} & = & \density \, \area \, \nablav \pot \\
\label{eq:uniform}
  \tens & = & \stress \, \area
\end{eqnarray}

In these equations $\s$ is a curvilinear coordinate along the tether, $\tensv$ is the
tension that the top part of the tether applies to the bottom part, $\pos$
is a position vector, $\area$ is the position dependent cross-sectional area of the tether,
$\density$ is the density of the tether, and $\stress$ is the stress in the
tether. Equation (\ref{eq:noshear}) expresses that the tether cannot bear any shear
load: the tension in the tether must be tangent to the cable. Equation
(\ref{eq:tens}) expresses Newton's second law: the change in tension along a
small piece of tether must oppose the force due to the potential.
Equation (\ref{eq:uniform}) expresses that the tether has a uniform
stress. Because the tether is uniformly stressed, we need not consider
elastic effects as they can be incorporated into $\stress$ and $\density$.

Equations (\ref{eq:tens}) and (\ref{eq:uniform}) can be combined to eliminate the area of
the cable.

\begin{equation}
\label{eq:tens-noa}
\der{\tensv}{\s} = \frac{\density \tens}{\stress} \nablav \pot 
\end{equation}

\subsection{Taper Profile}
\label{ssec:taper}

First we shall look at the tangential part of the static equations.
Integrating them will lead to a relation between the cable cross-section $\area$ and
the local potential $\pot$.

First, we take the dot product of (\ref{eq:tens-noa}) with $\diff{\pos}$,
divide by $\tens$, and use (\ref{eq:noshear}) to simplify.

\begin{equation}
\label{eq:taper}
\frac{\diff{\tens}}{\tens} = \frac{\density}{\stress} \, \diff{\pos} \cdot \nablav \pot 
\end{equation}

Integrating we get an expression for $\tens$ and therefore for $\area$.

\begin{equation}
\label{eq:taper-int}
\frac{\tens}{\tens_0} = \frac{\area}{\area_0} = e^{\frac{\density}{\stress}\pot}
\end{equation}

This formula shows that the area of the tether at a given position is
directly a function of the potential $\pot$ at that position. If $\deltaV$ 
is the difference in potential energy between the base of the tether
and the point where the potential energy reaches a maximum, then we can
express the taper ratio of the tether as:

\begin{equation}
\label{eq:taperratio}
\frac{\area_{max}}{\area_0}= e^{\frac{\density}{\stress}\deltaV}
\end{equation}

From this expression we can introduce a taper parameter
$\frac{\density}{\stress}\deltaV$ that characterizes the difficulty of building
a uniform-stress structure across a potential difference $\deltaV$. When it
is much smaller than one, almost no tapering is necessary. When it is much
larger than one the taper ratio becomes prohibitively large. This taper
parameter is closely related to the ratio of Pearson's characteristic
height~\cite{pearson75} to the geosynchronous altitude.

\subsection{Tether Shape Equation}

Projecting (\ref{eq:tens-noa}) onto a direction tangent to the tether allowed 
us to determine the taper profile. We now project perpendicularly to the
tether direction and combine with (\ref{eq:noshear}) to determine the shape 
the tether adopts in the gravitational potential:

\begin{equation}
\label{eq:shape}
\dder{\pos}{\s} = 
\der{\hat{\tens}}{\s} = \frac{\density}{\stress} \, \left( \nablav \pot \right)_\perp
\end{equation}

\noindent
where $\hat{\tens}=\tensv/\tens$ is a unit vector tangent to the
tether, and $(\nablav \pot)_\perp$ denotes the projection of
$\nablav \pot$ perpendicularly to $\hat{\tens}$.

Equation (\ref{eq:shape}) determines the tether's curvature. The tether
curves towards areas of higher potential, and the curvature is proportional
to the component of the gravity field that is normal to the tether.
This interpretation becomes more apparent in the case of a planar tether
where we can identify the direction of the tether by an angle $\phi$ so
that

\begin{equation}
\label{eq:phi-version}
\der{\phi}{\s} = \frac{\density}{\stress} \hat{e}_\phi \cdot \nablav \pot 
\end{equation}

\noindent where $\hat{e}_\phi$ is a unit vector in the direction of increasing~$\phi$.

\subsection{Boundary Conditions}

To have a complete description of the static tether, we additionally need to
consider boundary conditions. On one end, the tether is attached to the
anchor point on the planet. The anchor simply needs to provide a force 
equal to the tension in the tether
to keep the tether static. If the base of the tether isn't vertical
then there will be a horizontal component to this force, so the tether will
tend to pull the anchor sideways (see Section~\ref{sec:side-force}).

From equation (\ref{eq:taper-int}) we know that the tension in the tether
never goes to zero. Therefore, the free
end of the cable must have a force applied to it to balance the tension in
the cable. That force is provided by a counterweight of mass $m$ which must
satisfy:

\begin{equation}
\label{eq:boundary}
\tensv = -m \nablav \pot
\end{equation}

\noindent Thus the counterweight must be located in such a way that the
tether is tangent to the local gravity field. 

\section{The Rotating Coulomb Potential}
\label{sec:coulomb}

So far we have considered a uniform stress tether in an arbitrary potential
$\pot$.  To make further progress, we will now consider the specific
potential that applies in the case of the space elevator attached to a
planet. Because we are considering the statics of the tether, we have to
place ourselves in a reference frame that is rotating with the planet. Thus
the potential we are interested in is a combination of the Coulomb potential
of the planet's gravity and the centrifugal potential due to the planet's
rotation around its axis:

\begin{equation}
\label{eq:coulomb-pot}
\pot = -\frac{G M_p}{r} - \frac{1}{2} r_\perp^2 \Omega^2
\end{equation}

In this equation $G$ is the gravitational constant, $M_p$ is the mass of
the planet, $\Omega$ is the angular velocity of the planet's rotation, $r$
is the distance to the center of the planet, and $r_\perp$ is the distance
to the axis of rotation of the planet.

\subsection{Planar Tether Profile}

One of the first things we note about the potential is that it is invariant
by reflection about planes that contain the planetary axis of rotation.
This invariance must also apply to the resulting acceleration field.  Thus,
the forces caused by the potential will all be in a plane containing the
axis of rotation and the point at which they are applied.

Therefore, if we consider a plane containing the axis of rotation of the
planet and the counterweight, we find that the tension in the tether at the
counterweight is in that plane. As we move down the tether, the forces
acting on the tether are in that plane, so the tether remains in that plane
all the way to the anchor.

We conclude that the shape of the space elevator will be planar, even in
the non-equatorial case. This greatly simplifies the problem to be solved,
as we can now work in two dimensions in a plane that is perpendicular to the
equatorial plane.

\subsection{Non-Dimensional Problem}

Reducing a problem to non dimensional form is an excellent way of
extracting the physically important parameters. We now apply this
methodology to the space elevator.

First we note that the potential can be written in terms of the synchronous
radius $r_s=(G M_p / \Omega^2)^{1/3}$ and the characteristic potential 
$V_0 = (G M_p \Omega)^{2/3}$ in the form:

\begin{equation}
\label{eq:coulomb-nondim}
\pot = -\pot_0 \left( \frac{r_s}{r} + 
\frac{1}{2} \frac{r_\perp^2}{r_s^2} \right)
\end{equation}

Thus, $r_s$, the distance from the center of the planet to the synchronous
orbit, is the natural distance scale for this problem.
We shall therefore rewrite (\ref{eq:shape}) replacing all
distances $d$ by normalized distances $\tilde{d}=d/r_s$, and inserting the expression
for $V$ from~(\ref{eq:coulomb-nondim}):

\begin{equation}
\label{eq:shape-nondim}
\dder{\vec{\tilde{r}}}{\tilde{\s}} = \frac{\density \pot_0}{\stress} 
\left(\frac{\vec{\tilde{r}}}{\tilde{r}^3} - \vec{\tilde{r}}_\perp \right)_\perp
=-\alpha \left(\vec{\tilde{g}}\left(\vec{\tilde{r}}\right)\right)_\perp
\end{equation}

This is the differential equation that determines the shape of the tether
in a rotating Coulomb potential. This equation contains a single scalar
parameter $\alpha$ which we shall call the shape parameter.

\begin{equation}
\alpha = \frac{\density \pot_0}{\stress} 
\end{equation}

\noindent The shape parameter is the ratio of the characteristic potential
of the potential field to the strength to weight ratio of the tether
material. It also naturally appears in (\ref{eq:taper}),
(\ref{eq:taper-int}) and (\ref{eq:taperratio}) when they are applied to the
rotating Coulomb potential. The shape parameter
determines how deep it is possible to go into the normalized potential well before the
taper ratio becomes excessively high. 
Indeed, well below the synchronous
altitude, $\deltaV \approx \pot_0 r_s / r$, so the taper parameter is
approximately $\alpha / \tilde{r}$. Thus for $\alpha \ll \tilde{r}$, the
taper ratio is close to 1, while for $\alpha \gg \tilde{r}$, the taper
ratio is gigantic.

In the case of the Earth and the tether parameters from~\cite{edwards2002}, 
$\alpha \approx 0.189$ and $\tilde{r} \approx 0.151$. Thus we are close to
the limit of feasibility.

\subsection{Solving the Shape Equation Qualitatively}
\label{sec:qualitative}

To get an idea of the solutions of (\ref{eq:shape-nondim}) that satisfy the
boundary condition (\ref{eq:boundary}), it is useful to
study Figure~\ref{fig:field} which is a plot of the vector field
$\vec{\tilde{g}}$. We shall assume without loss of generality that the
anchor is in the upper right-hand quadrant ($x>0$ and $y>0$).  More
complete derivations can be found in~\cite{beletsky93}.

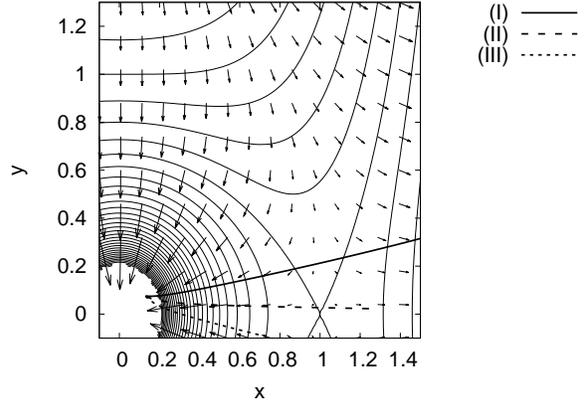
\mygraphfigure{fig:field}{width=\columnwidth}{fieldplot}{The normalized
rotating Coulomb field $\vec{\tilde{g}}$. The equatorial plane is horizontal, and
the North-South direction is vertical. Equipotential lines and field
values are plotted along with three example tether solutions. The tether
solutions are for $\alpha=0.189$, $\tilde{r_0}=0.151$, $\theta_0=30$\textdegree\ and
inclinations $\psi_0$ of 35\textdegree, 55\textdegree\ and 75\textdegree.}

To begin, we note that the North-South component of the field always points
towards the equator. This has two consequences. First, because of
(\ref{eq:tens-noa}) the $y$ component of $\tensv$ satisfies:

\begin{equation}
y \der{\tens_y}{s} > 0
\end{equation}

\noindent Second, because of (\ref{eq:boundary}), the tip of the tether at
the counterweight has to be sloped towards the equatorial plane (i.e., $y
\tens_y < 0$. Combining these two facts, we find that $\tens_y$ is negative
over the whole length of the tether. This implies via~(\ref{eq:noshear})
that the distance from the tether to the equatorial plane must
monotonically decrease as we move along the tether from the anchor point to
the counterweight. If the tether ever crosses the equatorial plane, or
starts heading away from the equatorial plane, the boundary condition will
never be satisfied.

Below the synchronous altitude, $\vec{\tilde{g}}$ is pointing down and to
the left. As we have seen, $T$ is pointing down and to the right. Therefore
because of~(\ref{eq:shape-nondim}) the tether only curves away from the
equator ($\phi$ increases monotonically). We will use this result in
Section~\ref{sec:small-planet}. 

\mygraphfigure{fig:lengthvsinclination}{width=\columnwidth}{lengthvsinclination}
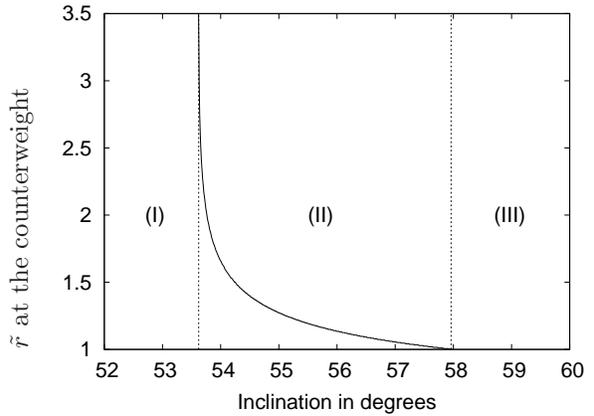
{Normalized distance from
the center of the Earth to the counterweight at latitude $\theta_0=30$\textdegree.}

Figure~\ref{fig:field} shows how solutions of~(\ref{eq:shape-nondim}) depend on the
inclination of the tether at the anchor. Case (II) satisfies the boundary
condition at the counterweight, while (I) and (III) extend infinitely
without the boundary condition ever being satisfied.

Indeed, if the inclination is too low as
in case (I), then the tether curves away from the equatorial plane before
suitable boundary conditions for a counterweight can be found. If the inclination
is increased, the point at which the tether is parallel to the equatorial
plane moves out towards infinity, and ceases to exist. At that point, it
becomes possible to satisfy the boundary condition at infinity. As the
inclination is increased more, the altitude of the counterweight gets lower and
lower, as in case (II). Once the altitude of the counterweight reaches 
the synchronous altitude, 
satisfying the boundary condition becomes impossible once again. The tether
crosses the equatorial plane before reaching the synchronous altitude
preventing it from being terminated as in case (III). 

We conclude that for a given anchor location, there will generally be a
range of inclinations for which a tether shape that satisfies the
counterweight boundary condition exists. Within that range, the
counterweight altitude decreases from infinity to the synchronous altitude
as in Figure~\ref{fig:lengthvsinclination}.

Figure~\ref{fig:inclination-vs-latitude} shows the inclinations that lead
to valid tether solutions in the case of the Earth space elevator. The
graph has been truncated at an inclination of 90\textdegree. Higher
inclinations are mathematically possible, but the tether would have to go
underground for the first part of its length.

\mygraphfigure{fig:inclination-vs-latitude}{width=\columnwidth}{inclinationrange}
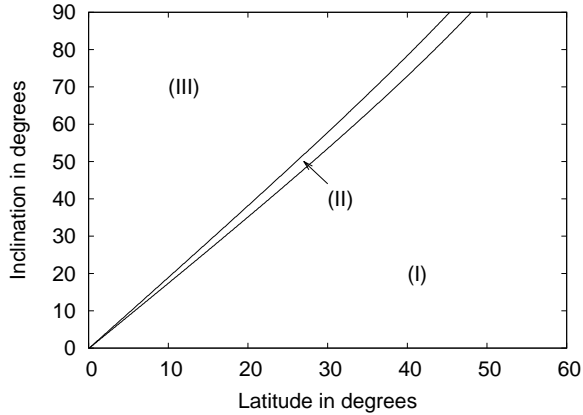
{Possible inclinations for an Earth space elevator as a function of
latitude ($\alpha=0.189$).}

\section{Maximum Anchor Latitude}
\label{sec:maxlatt}

As we saw in Figure~\ref{fig:inclination-vs-latitude}, there is a maximum
latitude beyond which no inclination allows the tether to satisfy the
counterweight boundary conditions. With Figure~\ref{fig:maxlatitude-90}, it is 
possible to determine the maximum anchor latitude in the general case.
This figure was generated by considering, for a given planetary radius
and shape parameter, which latitude would lead to an infinitely long tether
with 90\textdegree\ inclination.

\myfigure{fig:maxlatitude}
{

\mygraphsubfigure{fig:maxlatitude-90}{width=\columnwidth}{maxlattitude-90}
{Maximum inclination of 90\textdegree.}

\mygraphsubfigure{fig:maxlatitude-45}{width=\columnwidth}{maxlattitude-45}
{Maximum inclination of 45\textdegree.}

}
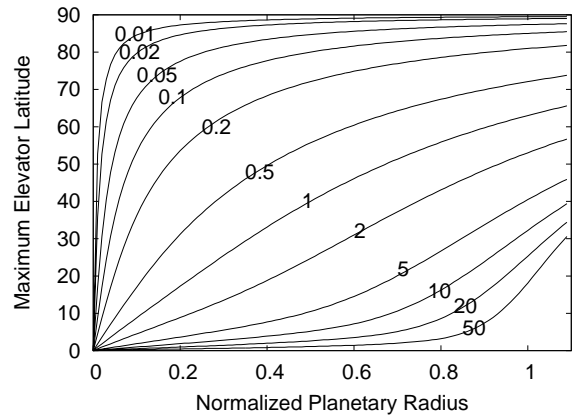
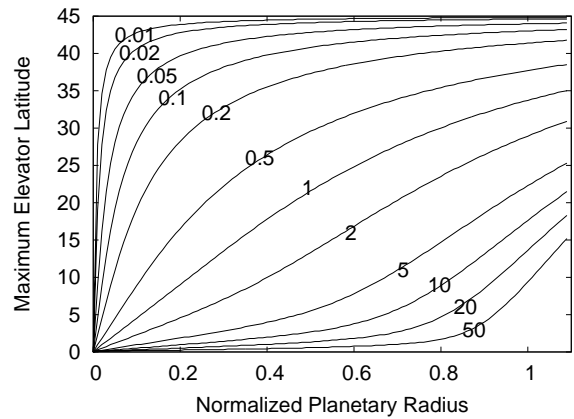
{Maximum reachable latitude as a function of normalized planetary radius
for different values of the shape parameter $\alpha$, assuming maximum
tether inclinations at the anchor of 90\textdegree\ and 45\textdegree.}

This figure isn't very useful for practical purposes because it accepts
tethers that are very inclined at the base. As we shall see in
Section~\ref{sec:cost-of-inclination}, such tethers have a small payload.
Therefore, we have also considered the case where tether
inclination is limited to 45\textdegree\ in
Figure~\ref{fig:maxlatitude-45}.

Clearly these two figures are very similar except for a different scaling
of the vertical axis. By considering a number of limit cases we shall try
to explain this similarity and better understand the characteristics of
the plots. For clarity we introduce the function 
$\theta_{max}(\tilde{r}_0, \psi_{max}, \alpha)$ which gives the maximum
latitude from the normalized planetary radius, the maximum acceptable tether
inclination and the shape parameter. 

$\theta_{max}$ is the latitude at which a tether inclined by $\theta_{max}$
at the anchor is right at the limit between case (I) and case (II). At that
limit, the tether solution is infinitely long. So, to study $\theta_{max}$, we shall be
studying tether solutions that go to infinity.

\subsection{Small Planet Limit}
\label{sec:small-planet}

First we shall direct our attention to the case where $\tilde{r}_0 \ll 1$.
This approximation corresponds to the case where the planet is small
compared to the synchronous altitude (we could also say that the planet
rotates slowly compared with an object that would orbit near ground level).
This is a good approximation for most known planetary bodies; the gas
giants are the main exception. For Earth $\tilde{r}_0 \approx 0.151$ and
for Mars $\tilde{r}_0 \approx 0.166$. 

As always (see Section~\ref{sec:qualitative}), the angle $\phi$ decreases 
with altitude, as does the distance to the equatorial plane. Because the 
planet is small, the distance to the equatorial plane at the anchor is much
smaller than the distance to the synchronous altitude. Therefore, to avoid
crossing the equatorial plane, the tether is nearly
parallel to the equatorial plane far before the synchronous altitude. This means
that any significant tether curvature occurs well below that altitude.
Well below synchronous altitude, the centrifugal force term
in $\vec{\tilde{g}}(\vec{\tilde{r}})$ can be ignored, so
(\ref{eq:shape-nondim}) reduces to

\begin{equation}
\label{eq:shape-nondim-r-small}
\dder{\vec{\tilde{r}}}{\tilde{\s}} = \alpha 
\left(\frac{\vec{\tilde{r}}}{\tilde{r}^3}\right)_\perp
\end{equation}

This equation can be normalized by changing the length scale by a factor
$\alpha$

\begin{equation}
\label{eq:shape-nondim-r-small-renorm}
\dder{\vec{\breve{r}}}{\breve{\s}} =
\left(\frac{\vec{\breve{r}}}{\breve{r}^3}\right)_\perp
\end{equation}

\noindent where the distance $\breve{d}$ is the normalized version of
$\tilde{d}$ and corresponds to $\tilde{d}/\alpha$.
This new normalization is very satisfying as the tether equation contains
no constants at all. However, to prevent confusion with the previous normalization of
distances, we will avoid this notation whenever possible.

The consequence is that $\theta_{max}$ is only a function of  
$\tilde{r}_0/\alpha$ and $\psi_{max})$. There is one
fewer parameters to consider. In Figure~\ref{fig:maxlatitude}, for small
values of the normalized planetary radius, the curves for different shape
can be deduced from one another by stretching in the horizontal direction.

\subsection{Low Curvature Limit}

Remaining in the small planet limit, we now consider the case where
$\tilde{r}_0/\alpha \gg 1$. In this case, the tether undergoes very little
curvature at all. This is particularly clear 
from~(\ref{eq:shape-nondim-r-small-renorm}) where there is very little
effect on the tether when $\breve{r}$ is large.

If we ignore curvature altogether, we find that the tether is straight. In this
approximation, $\theta_{max}\approx \psi_{max}$. 

%

\subsection{High Curvature Limit}

Still in the small planet limit, we now consider the opposite case in
which $\tilde{r}_0/\alpha \ll 1$. In this case, the tether curves
sharply if its inclination $\psi$ is not small. If the tether is
inclined at its base, it very quickly becomes vertical. This prevents large
starting latitudes.

Since the latitude is small and the curvature occurs near the base of the
tether, we will make the approximation of a uniform gravity field over a flat planet. 
In this approximation $\phi \approx -\psi$. We normalize equation~(\ref{eq:phi-version})
and apply the small planet limit to get

\begin{equation}
\label{eq:shape-small-latt}
\der{\phi}{\tilde{\s}} = -\frac{\alpha}{\tilde{r}_0^2} \sin(\phi)
\end{equation}

\noindent which can be further simplified by taking the derivative with
respect to $\tilde{y}$ instead of $\tilde{s}$

\begin{equation}
\label{eq:shape-small-latt2}
\der{\phi}{\tilde{y}} = -\frac{\alpha}{\tilde{r}_0^2}
\end{equation}

We now integrate from 0 to $y_0$, and note that $\phi=0$ at $y=0$,
to get an expression for the inclination at the anchor

\begin{equation}
\psi_0=-\phi_0=\frac{\alpha}{\tilde{r}_0^2} y_0
\end{equation}

Finally, we can express $y_0$ in terms of $r_0$ and the starting latitude
$\theta_0$ as $y_0\approx r_0 \theta_0$ to get

\begin{equation}
\psi_0=\frac{\alpha}{\tilde{r}_0} \theta_0
\end{equation}

So in the high curvature limit

\begin{equation}
\theta_{max}\approx \frac{\tilde{r}_0}{\alpha} \psi_{max}
\end{equation}

\subsection{A Combined Formula}

If $\tilde{r}_0/\alpha$ is near 1 then the analysis becomes much more
difficult as both the $x$ and $y$ gravity components are significant. A
simple empirical interpolation can nevertheless be used with very good
results

\begin{equation}
\label{eq:atan}
\theta_{max}\approx \frac{2}{\pi} \psi_{max} \arctan\left(\frac{\pi}{2}
\frac{\tilde{r}_0}{\alpha}\right)
\end{equation}

It is easy to verify that this formula holds in both limit cases.
When $\tilde{r}_0/\alpha$ is near 1, this formula gives a result that is up
to 8\% too high. 

\mygraphfigure{fig:maxlatitude-atan}{width=\columnwidth}{maxlattitude-atan}
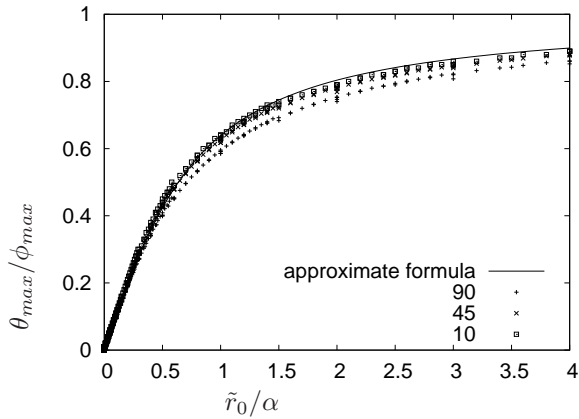
{Comparison of~(\ref{eq:atan}) with simulation data for maximum
inclinations of 10\textdegree, 45\textdegree\ and 90\textdegree. Only data
points with $\tilde{r}_0<0.3$ were plotted.}

Figure~\ref{fig:maxlatitude-atan} illustrates the quality
of the approximation. The match is slightly better for low values of the
tether inclination. For high inclinations the approximation is slightly
optimistic. In this figure we have limited ourselves to
$\tilde{r}_0<0.3$, for $\tilde{r}_0>0.5$ we start to see significant
deviation from~(\ref{eq:atan}). With larger $\tilde{r}_0$ higher latitudes
than expected can be reached.


\section{Practical Considerations}
\label{sec:practical}

So far we have considered whether it was possible to have a space elevator
at a given latitude by considering the static tether equation. In practice
other considerations will further limit the latitude that can be reached.

\subsection{Payload to Elevator Mass Ratio}
\label{sec:cost-of-inclination}

One of the major concerns in space elevator construction is the ratio of
payload mass to elevator mass. Indeed, this ratio determines how much
material has to be lifted during the elevator construction for a given
elevator capacity.

We saw in Section~\ref{ssec:taper} that the taper ratio of a uniform-stress 
tether only depends on the change in potential along the tether. The
potential is uniform at the surface of a planet, and from
Figure~\ref{fig:field} the potential changes very slowly near the
synchronous orbit. Therefore, the taper ratio for non-equatorial space
elevators is almost the same as the taper ratio for equatorial space
elevators.

In the small planet limit, the angle between the tether and the equatorial
plane is small, except possibly near the surface of the planet. Therefore,
the length of the tether doesn't depend much on the latitude of the
anchor. Moreover, since the potential depends slowly on $y$, the taper
profile of the non equatorial space elevator is nearly the same as the
profile for an equatorial one.

Therefore, the only significant difference between equatorial and
non-equatorial elevators is due to the tension at the 
base of the tether not being vertical in the non-equatorial case. Since only the
vertical component of the tension can be used to lift a payload, and
elevators of equal mass have nearly equal tension at their base, we get
a reduced payload to elevator mass ratio in the non-equatorial case:
\begin{equation}
\left(P/M\right)_{off-equator}\approx 
\left(P/M\right)_{equator} \cos(\psi_0)
\end{equation}

Thus to maintain payload when leaving the equator means one has to multiply
the elevator mass by $1/\cos(\psi_0)$. For small inclinations at the anchor
this inefficiency is insignificant. But approaching inclinations of
90\textdegree\ is clearly out of the question.

The designer of Earth space elevators will find Figure~\ref{fig:lattvssigma} 
useful to quickly determine the maximum elevator latitude as a function of
maximum inclination at the anchor. For ease of use $\sigma_0$ has been
used instead of the shape parameter, the tether density is assumed to be
fixed at 1300 kg/m$^3$.

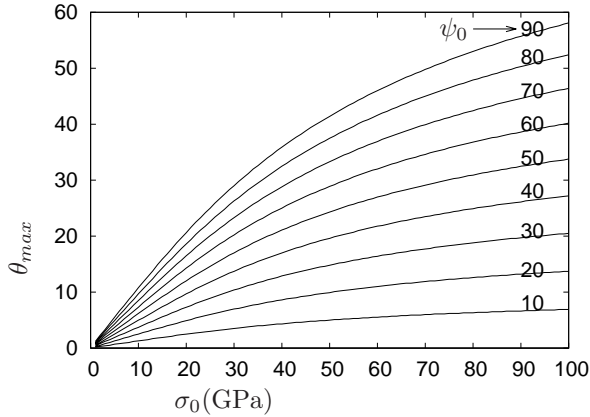
\mygraphfigure{fig:lattvssigma}{width=\columnwidth}{lattvssigma}{Maximum
reachable latitude for an Earth space elevator as a function of 
$\sigma_0$, for different values of the inclination $\psi_0$.}

\subsection{Horizontal Force on Anchor}
\label{sec:side-force}

In addition to reducing the payload to mass ratio of the elevator, the inclination 
at the tether's base causes a horizontal force to be applied to the anchor
platform. This force is simply given by:
\begin{equation}
F=T_0 \tan(\psi_0)
\end{equation}

If the anchor is a mobile ocean platform, this force will need to be
countered or else the anchor will drift towards the equator. For heavy lift
elevators significantly off the equator, this force will have to be taken
into account when selecting the anchor platform.

\subsection{Stability}

An equatorial tether with very high elasticity (low Young's modulus), or
with a small extension beyond the synchronous altitude can be unstable
\cite{beletsky93}. For equatorial space elevators, we would be considering
conditions far from this instability region, so it is of no concern. In the
case of non-equatorial elevators, the curvature at the base of the elevator
should cause a reduction in the axial stiffness of the tether as seen from
the counterweight. We can therefore conjecture that the frequency of the
elevator modes will decrease for a non-equatorial elevator. This could
cause the instability to occur for realistic tether parameters.

We conjecture that as in the equatorial case, the instability will occur
for short tethers, near the boundary between (II) and (III), where the
counterweight is just beyond the synchronous altitude (see
Section~\ref{sec:qualitative}). However the instability will extend to
greater counterweight altitudes. The maximum reachable is determined at the
(I)-(II) boundary and should not be affected. However, external effects
such as the presence of Earth's Moon may limit the length of the elevator,
and thus limit the latitude.

\subsection{Deployment}

During the initial deployment phase, there is no contact between the tether
and the Earth, which takes away the only force keeping the elevator away
from the equatorial plane. This leaves two possibilities for deploying a non-equatorial
space-elevator. 

\begin{itemize}

\item A propulsion system can be attached to the bottom of the tether
during deployment to keep the tether away from the equatorial plane. For a
10\textdegree\ latitude, and a one ton elevator this option would require
hundreds of newtons of thrust over a period of days and is therefore
impractical.

\item The tether can be deployed in the equatorial plane,
attached to the anchor platform, and then moved to its final location away
from the equator. If an off-equator location has been selected to avoid
interfering with geosynchronous satellites, the initial deployment can be
done at a longitude where there is an available geosynchronous slot, after
which the elevator can be moved to its final latitude.

\end{itemize}

\subsection{Tether Shape Determined Numerically}

Three of the applications we mentioned for non-equatorial space
elevators were the avoidance of particular areas in the equatorial plane.
In this paper we have
not pushed the analysis of~(\ref{eq:shape-nondim}) far enough to determine
whether these obstacles are indeed avoided. Figure~\ref{fig:shapes} shows
some numerical solutions. They suggest that avoiding the geosynchronous
satellites is easy. On the other hand, the most intense areas of the 
radiation belts extend over 2,000~km above the equatorial plane, which only
highly inclined elevators can avoid.

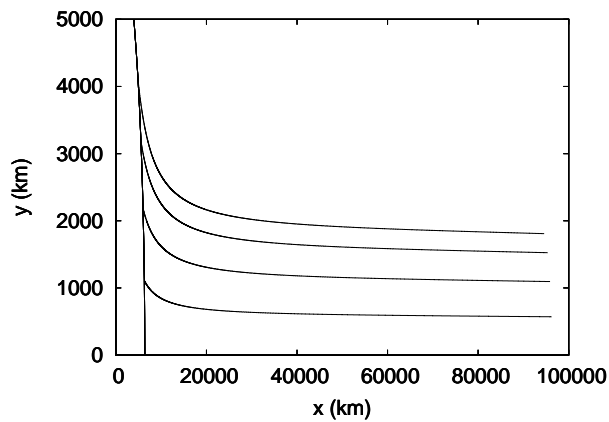
\mygraphfigure{fig:shapes}{width=\columnwidth}{shape}{Numerical solutions
of the shape equation for the Earth with the standard tether parameters.
The length of the tether was set to 90,000~km. Starting latitudes of
10\textdegree, 20\textdegree, 30\textdegree\ and 40\textdegree. Note that the scale is
different along the $x$ and $y$ axes.}

\section*{Conclusion}

In this paper we have presented the equations that govern the statics of
non-equatorial uniform-stress space elevators. These equations have been
reduced to a non-dimensional form allowing the analysis to be applied to
any tether and planetary parameters.

The tether's taper profile has turned out to be easy to compute as in
the equatorial case, once its the spatial configuration is known.
Unfortunately, the spatial configuration is difficult to obtain
analytically.

Of particular interest to the elevator designer is the maximum
anchor latitude for a non-equatorial elevator. This problem
has been solved in a few limit cases, and an approximate formula has been 
proposed.

Off the equator, the tether is not vertical at the base of the elevator.
This causes a reduction in payload, which is the major engineering cost of being
off the equator. It also causes a horizontal force to be applied to the
anchor station, which much be taken into account for an ocean platform.

This study has ignored dynamic effects, in particular the
stability of off-equatorial elevators has to be checked. For small
latitudes the stability of equatorial elevators carries over, but we
expect instabilities to occur in a wider range of conditions than in the
equatorial case. This remains to be verified. The effect of climbers on the
tether also needs to be studied.

\section*{Acknowledgements}
I would like to thank Val\'erie Leblanc for the time she spent proof reading
and pointing out bugs.

\bibliographystyle{IEEEtran}
\bibliography{paper}

\end{document}